\documentclass[twocolumn, prl, showpacs, preprintnumbers,amsmath,amssymb, superscriptaddress]{revtex4}
\usepackage{graphicx}
\usepackage{dcolumn}
\usepackage{bm}
\usepackage{color,ulem} 

\begin{document}

\title{Light-emitting waveguide-plasmon polaritons}

\author{S. R. K. Rodriguez} \email{s.rodriguez@amolf.nl}
\affiliation{Center for Nanophotonics, FOM Institute AMOLF, c/o
Philips Research Laboratories, High Tech Campus 4, 5656 AE
Eindhoven, The Netherlands}

\author{S. Murai}  \email{murai@dipole7.kuic.kyoto-u.ac.jp}
\affiliation{Center for Nanophotonics, FOM Institute AMOLF, c/o
Philips Research Laboratories, High Tech Campus 4, 5656 AE
Eindhoven, The Netherlands} \affiliation{Department of Material
Chemistry, Graduate School of Engineering, Kyoto University,
Katsura, Nishikyo-ku, Kyoto 615-8510, Japan }

\author{M. A. Verschuuren}
\affiliation{ Philips Research Laboratories, High Tech Campus 4,
5656 AE Eindhoven, The Netherlands.}

\author{J. G\'{o}mez Rivas}
\affiliation{Center for Nanophotonics, FOM Institute AMOLF, c/o
Philips Research Laboratories, High Tech Campus 4, 5656 AE
Eindhoven, The Netherlands} \affiliation{COBRA Research Institute,
Eindhoven University of Technology, P.O. Box 513, 5600 MB Eindhoven,
The Netherlands}

  \date{\today}

\begin{abstract}
We demonstrate the generation of light in an  optical waveguide
strongly coupled to a periodic array of metallic nanoantennas. This
coupling gives rise to hybrid waveguide-plasmon polaritons (WPPs),
which undergo a transmutation from plasmon to waveguide mode and
viceversa as the eigenfrequency detuning of the bare states transits
through zero.  Near zero detuning, the structure is nearly
transparent in the far-field but sustains strong local field
enhancements inside the waveguide. Consequently, light-emitting WPPs
are strongly enhanced at energies and in-plane momenta for which
WPPs minimize light extinction. We elucidate the unusual properties
of these polaritons through a classical model of coupled harmonic
oscillators.

\end{abstract}

\maketitle

The field of plasmonic metamaterials has inherited an invaluable
legacy from atomic physics for the interpretation of non-trivial
spectral line shapes. Seminal examples are Fano's theory  on
asymmetric resonance line shapes~\cite{Fano}, and
Electromagnetically Induced Transparency (EIT)~\cite{Harris91}. Fano
described the quantum interference between a continuum of states and
a discrete state whose energy lies within the continuum. His theory
has found broad applicability  to classical systems
also~\cite{FanoRev10, FanoNat}, where similar line shapes arise from
the interaction of spectrally broad and narrow resonances. The
connected phenomenon of EIT, wherein destructive quantum
interference between different excitation pathways renders a
spectrally narrow transparency window within an absorbtion band of
an atomic medium~\cite{Marangos, Fleis}, has also been a major
inspiration for the metamaterials community~\cite{Zhang08,
Giessen09, Yannopapas09, Soukoulis09, Kekatpure&Brongersma10,
Zhang11APL}. The abrupt changes in frequency dispersion associated
with EIT enable light to be slowed~\cite{Hau99}, or even
halted~\cite{Hau01}. Despite the fundamental distinction in the
nature of quantum versus classical waves, i.e., probability
amplitudes rather than classical field amplitudes interfere, Fano,
EIT, and even tunneling-like behavior~\cite{Woerdman, Novotny10a}
manifest for both due to the ubiquity of wave phenomena.

In this Letter, we demonstrate the generation of light in a strongly
coupled classical system displaying Fano and EIT-like spectral line
shapes. We investigate the coupling of Localized Surface Plasmon
Polaritons (LSPPs), i.e., collective oscillations of conduction
electrons in metallic nanostructures, to a guided mode in a
light-emitting slab. LSPPs enable the  conversion of free space
radiation into localized energy and viceversa, empowering metallic
nanostructures as antennas for light~\cite{Hecht05, Novotny11}. When
a waveguide is placed near nanoantennas, the strong coupling of
LSPPs to guided modes leads to the formation of a new quasiparticle
in light extinction, known already as a Waveguide-Plasmon Polariton
(WPP)~\cite{Giessen03}. We herein report on a novel class of
light-emitting WPPs which display an extraordinarily enhanced
emission at frequencies of strong dispersion and nearly perfect
far-field transparency.  We support our experimental findings with
numerical simulations and a coupled oscillator model, explaining how
light emission can be resonantly enhanced in nearly transparent
optical media. Emission enhancements at energies and in-plane
momenta of far-field transparency hold great promise for plasmonic
based solid-state light-emitting devices with negligible absorption
losses. By reciprocity, a sensor can be cloaked such that its local
sensitivity is enhanced while remaining invisible to distant
observers~\cite{Alu09, Garcia09}.

\begin{figure}
\centerline{\scalebox{0.65}{\includegraphics{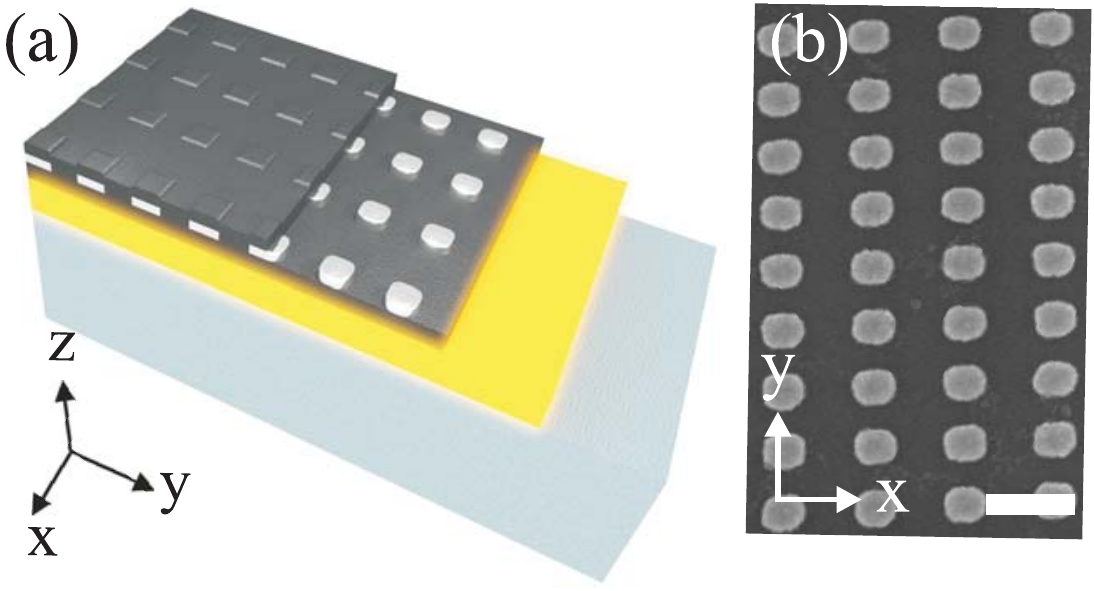}}}\caption{
(a) 3D view of the light-emitting waveguide structure coupled to a
silver nanoantenna array. The layers extend up to different
positions along the y-axis for illustration purposes only. From
bottom to top, the layers are, 1 mm of SiO$_2$, 230 nm of YAG:Ce,
and 20 nm silver nanoantennas surrounded by 20 nm of Si$_3$N$_4$ on
top and bottom. The light field emitted from the YAG:Ce  couples
localized surface plasmon polaritons in the silver nanoantennas to
guided photons in the YAG:Ce slab waveguide. (b) SEM image of the
silver nanoantenna array. The scale bar denotes 300 nm.}\label{fig1}
\end{figure}

Figure~\ref{fig1}(a) illustrates the structure we investigate. As a
light-emitting slab waveguide, a 230 nm layer of Yttrium Aluminum
Garnet doped with Ce$^{3+}$ ions (YAG:Ce) was fabricated by a
sol-gel method onto a fused silica substrate as described in
Ref.~\cite{Murai}. The $\sim$ 0.4 eV full width at half maximum
(FWHM) emission of YAG:Ce allows us to study light-emitting WPPs in
a wide spectral range. A 20 nm layer of Si$_3$N$_4$ was deposited on
top of the slab for a two-fold purpose: i) to planarize the surface,
and ii) to avoid emission quenching of Ce$^{3+}$ ions in proximity
to the metal~\cite{Wokaun83, Anger&Novotny06}. An Ag nanoantenna
array with a size of $2 \times 2$ mm$^2$ was fabricated by substrate
conformal imprint lithography~\cite{SCIL} onto the Si$_3$N$_4$
layer. Figure~\ref{fig1}(b) shows a scanning electron microscope
image of the array. The dimensions of the nanoantennas are $90
\times 70 \times 20$ nm$^3$, and the lattice constants are $a_x$ =
300 nm and $a_y$ = 200 nm. The array was covered by a conformal
layer of Si$_3$N$_4$ with a thickness of 20 nm to protect the Ag
nanoantennas from oxidation. We will show later (Fig.~\ref{fig4})
that, although Si$_3$N$_4$ ($n\approx 2.0$) has a higher refractive
index than YAG:Ce ($n \approx 1.7$, as determined from ellipsometry)
in the part of the visible spectrum where we work, the thinness of
the Si$_3$N$_4$ layer and the presence of the metallic antennas
enable the excitation of a guided mode in the YAG:Ce layer.

\begin{figure}
\centerline{\scalebox{0.65}{\includegraphics{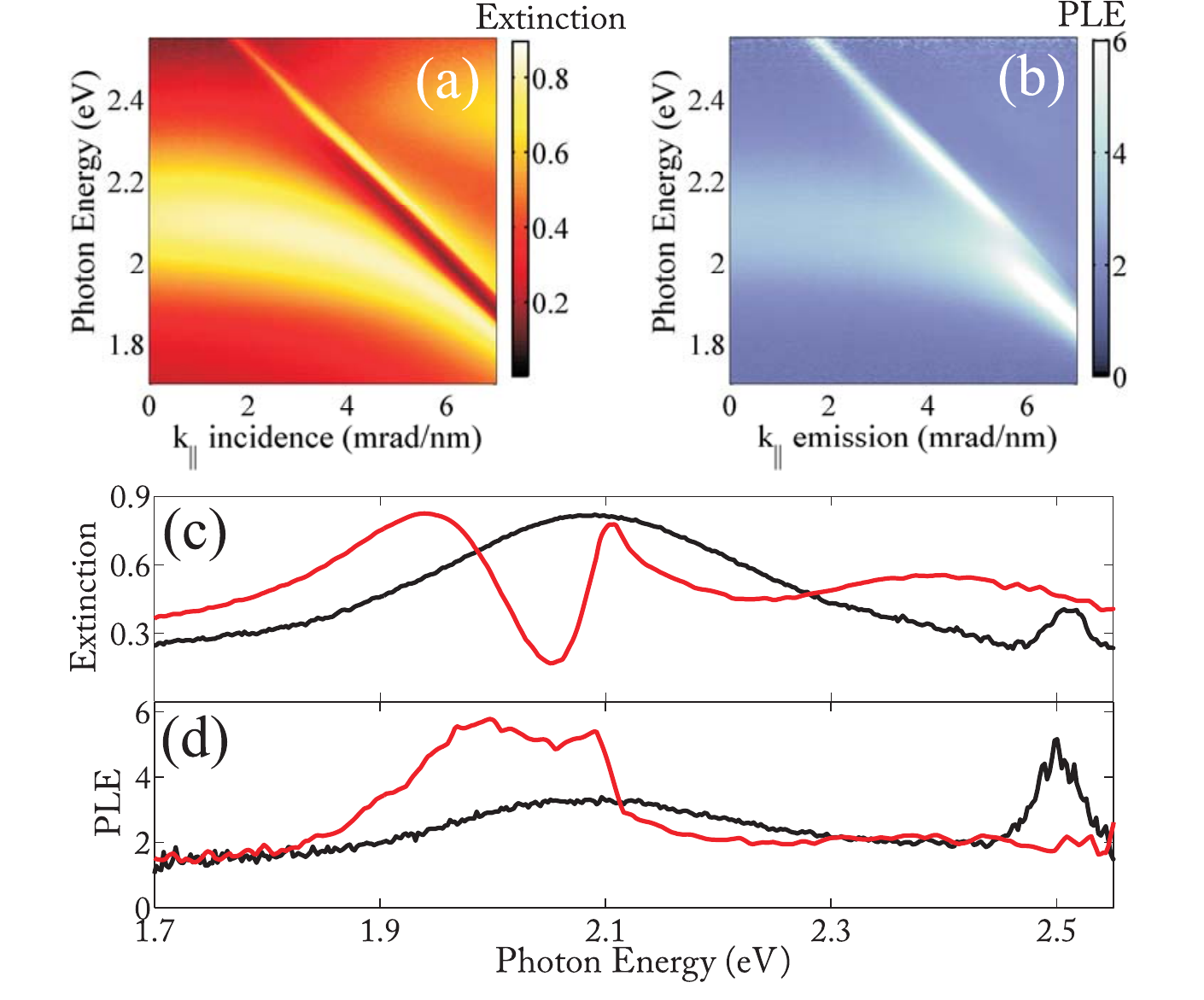}}} \caption{(a) Extinction, and (b) PhotoLuminescence Enhancement (PLE) of the emission from the YAG:Ce layer
coupled to the antennas, normalized to the layer without antennas. (c) and (d) are cuts of (a) and (b), respectively, at $k_{||}=2.1$ mrad/nm (black lines) and
$k_{||}=5.5$ mrad/nm (red lines).}\label{fig2}
\end{figure}

We performed variable angle, polarization-resolved, extinction
spectroscopy. The white light from a halogen lamp was collimated and
linearly polarized parallel to the short axis of the nanoantennas
(y-axis in Fig.~\ref{fig1}). The sample was rotated around the
y-axis, resulting in s-polarized incidence. The zeroth-order
transmittance $T_0$ was collected by a fiber-coupled
spectrophotometer. The extinction, defined as 1-$T_0$, is shown in
Fig.~\ref{fig2}(a) as a function of the incident photon energy and
the wave vector component parallel to the long axis of the antennas,
$\mathbf  k_\|= k_0 \sin(\theta_{in}) \hat{x}$, with $k_0$ the free
space wave vector and $\theta_{in}$ the angle of incidence. Computer
controlled rotation stages with an angular resolution $<0.2^\circ$
($k_{||} \approx 35$ $\mu$rad/nm at 2 eV) were used for the
measurements. We shall refer to the magnitude of $\mathbf k_{||}$ as
$k_{||}$.

The feature near 2.1 eV at normal incidence in Fig.~\ref{fig2}(a) is
associated with the dipolar LSPP along the short axis of the
antennas. Its flat angular dispersion and broad linewidth are
characteristic of localized resonances. The small size of the
particle along the polarization axis (70 nm) determines the
primarily dipolar response. The narrow feature on the high energy
side corresponds to the fundamental $TE$ guided mode in the YAG:Ce
slab. Its dispersion has been folded onto the first Brillouin zone
by the periodic nanoantenna array. As $k_{||}$ increases, the LSPP
acquires an angularly dispersive character and at  narrower
linewidth, whereas the opposite occurs for the guided mode. Thus,
their properties are interchanged.

To obtain the light emission dispersion diagram, the sample was
photoexcited by a continuous wave laser with an energy of 2.8 eV,
FWHM = 30 meV, a power far below saturation, impinging at an angle
of $10^\circ$. The photoluminescence passed through an analyzer, and
was collected as a function of the angle $\theta_{em}$ subtended by
the detector and the normal to the sample. The detector was rotated
around the y-axis, collecting s-polarized light with $\mathbf k_\|=
k_0 \sin(\theta_{em}) \hat{x}$.  The angular resolution was the same
as in the extinction measurements. The sample was excited
non-resonantly with regards to the plasmonic structure, as will be
shown next.

Figure~\ref{fig2}(b) shows the measured PhotoLuminescence
Enhnancement (PLE) , defined as  $I_{in}/I_{out}$ with $I_{in}$ and
$I_{out}$ the emission from the YAG:Ce slab in and out the presence
of the nanoantenna array, respectively. For small $ k_{||}$, where
the LSPP and guided mode resonances are largely detuned, the PLE
displays features qualitatively resembling those in extinction. Near
zero detuning, occurring for $k_{||} \approx 5.5$ mrad/nm, a
remarkable contrast between extinction and PLE takes place. We
illustrate examples of large detuning $k_{||}=2.1$ mrad/nm (black
line) and near zero detuning $k_{||}=5.5$ mrad/nm (red line), for
light extinction in Fig.~\ref{fig2}(c) and PLE in
Fig.~\ref{fig2}(d). For large detuning, the broad and strong
extinction feature associated with the LSPP leads to a modest PLE
with a similar line shape. The narrow and weak feature with a
Fano-like line shape associated with the guided mode  leads to a
larger PLE. This contrasting behavior has its origin in the spatial
field distributions associated with the LSPP and guided mode, which
we illustrate later by means of simulations. Near zero detuning, it
is not possible to discriminate between the LSPP and guided mode. In
this case, the extinction displays an EIT-like line shape with  a
nearly perfect transparency window. Remarkably, the emission from
the YAG:Ce waveguide is largely enhanced in this far-field
transparent and strongly dispersive spectral region, resulting in a
strong and spectrally broad light-emitting WPP. Notice that far from
any resonance the PLE tends to $1$. This indicates that pump
enhancements, i.e., resonant processes at the excitation energy, are
negligible. By matching the momentum of the pump photons to the
momentum of the waveguide photons with the assistance of the
grating, the magnitude of the PLE could be further boosted while
preserving the dispersive properties of the light-emitting WPPs.
Alternatively, the pump photons may be evanescently coupled into the
YAG:Ce waveguide as recently shown in Ref.~\cite{Murai}. Therein, a
30-fold pump enhancement was reported with negligible enhancement at
the emission energies.

\begin{figure}
\centerline{\scalebox{0.65}{\includegraphics{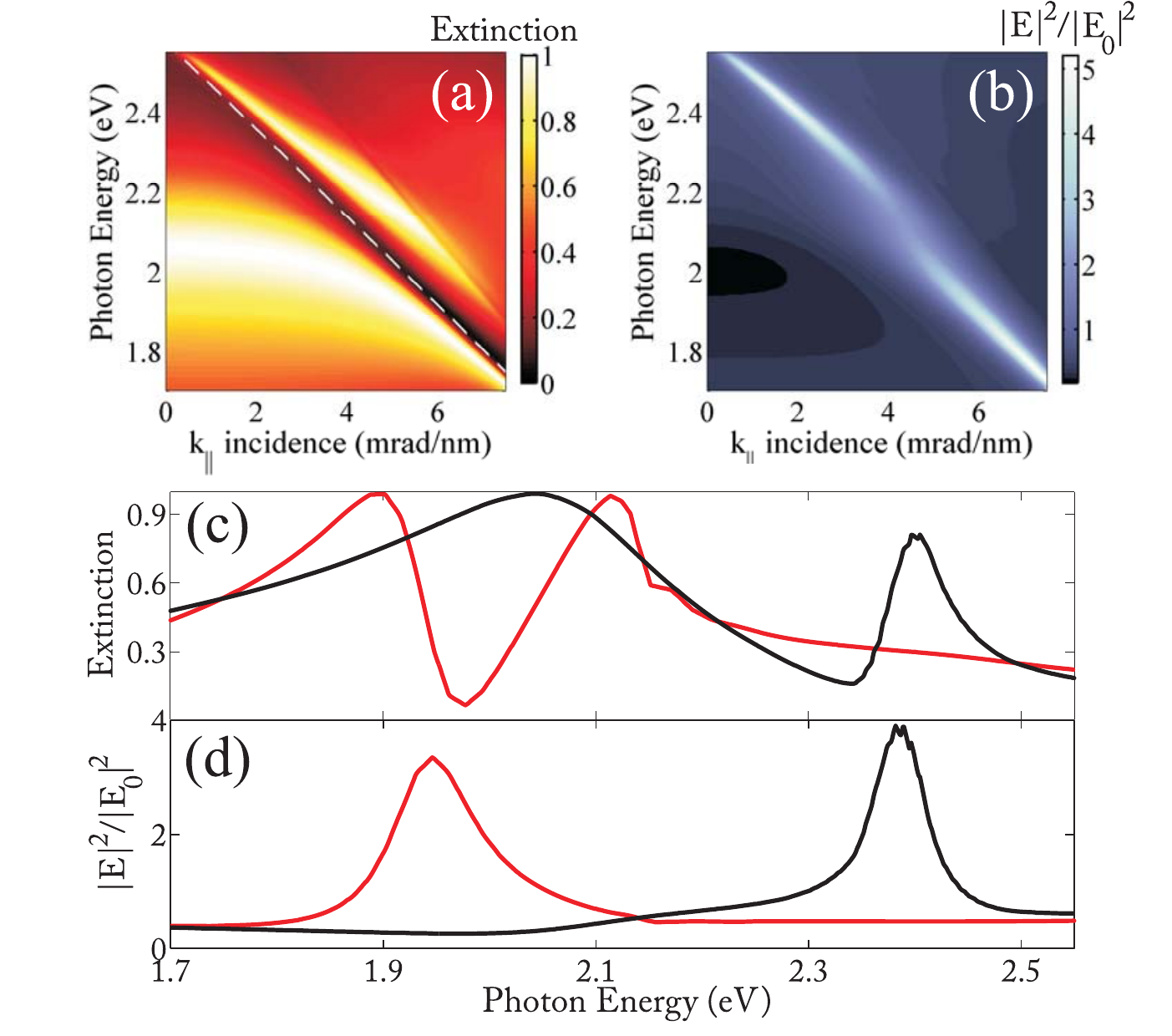}}} \caption{Finite element method simulations of (a) the extinction of an array of metallic nanowires with 70
nm width and 300 nm periodicity, and (b) the corresponding average field intensity enhancement in the YAG:Ce slab. (c) and (d) are cuts of (a) and (b),
respectively, at $k_{||}=2.1$ mrad/nm (black lines) and $k_{||}=5.5$ mrad/nm (red lines).}\label{fig3}
\end{figure}

 We performed finite element method simulations (Comsol) under plane wave illumination to
illustrate the field distributions associated with the coupled
optical modes. For ease of computation, we chose a 2D system with
the same periodicity as  the experimental structure along the x-axis
(the direction of the in-plane momentum). The width and height of
the simulated metallic wire were chosen to be 70 nm and 20 nm,
respectively. These dimensions determine the spectral features of
the LSPP. For $TM$ polarization, the coupling between the LSPPs
along the width of the metallic wire and the $TM$ guided mode gives
rise to the same physics in light extinction observed in our 3D
system for $TE$ polarization~\cite{Giessen03}. The optical data for
Ag was taken from Ref.~\cite{Palik91}, for YAG:Ce it was taken from
ellipsometry measurements not shown here, and we set a constant
refractive index of $n=2.0$ for Si$_3$N$_4$ and $n=1.43$ for
SiO$_2$. Floquet boundary conditions were used for the sides of the
unit cell, and Perfectly Matched Layers (PMLs) on the top and
bottom.

Figure~\ref{fig3}(a) shows the extinction obtained from the
simulations. The white dashed line indicates the dispersion relation
of the fundamental $TM$ guided mode in a three-layer structure,
calculated as outlined in Ref.~\cite{Yariv}. The structure consists
of SiO$_2$ with n=1.43, an effective YAG:Ce + Si$_3$N$_4$ slab
assuming their average refractive index weighted by the
corresponding fractional area, and air with n=1. The good agreement
between the white line and the transparency band in
Fig.~\ref{fig3}(a), arising at the eigenvalue of the (bare) guided
mode, justifies the assumption for determining the refractive index
of the middle layer. Figure~\ref{fig3}(b) shows the simulated
electric field intensity enhancement in the YAG:Ce slab, defined as
$|E|^2 / |E_0|^2$ with $E$ and $E_0$ the total and incident electric
field, respectively, both spatially averaged over the interior of
the YAG:Ce slab. In Figs.~\ref{fig3}(c) and ~\ref{fig3}(d) we make
cuts of the extinction and $|E|^2 / |E_0|^2$, respectively, at a
large detuning of $k_{||}=2.1$ mrad/nm (black lines) and near zero
detuning of $k_{||}=5.5$ mrad/nm (red lines). At a large detuning we
observe that  $|E|^2 / |E_0|^2$ is significantly greater at the
guided mode energy (2.4 eV) than at the LSPP energy (2.05 eV). The
large electromagnetic enhancements occurring in the locality of the
emitters explain the greater emission enhancement measured for the
guided mode at large detunings. Near zero detuning, $|E|^2 /
|E_0|^2$ displays a maximum at energies within the far-field
transparency band. This counterintuitive behavior arises from the
interference nature of the extinction, which allows for a partial
cancelation of the far-field response of a system sustaining intense
local fields. This property of extinction is at the heart of recent
theoretical proposals for cloaking a sensor~\cite{Alu09, Garcia09}.
It transpires that near zero detuning the LSPP-guided-mode coupling
induces a transparency band through far-field interference, yet the
emission from the Ce$^{3+}$ ions is enhanced due to  the local field
enhancements shown in Fig.~\ref{fig3}.

 We present in Fig.~\ref{fig4} the total electric field  normalized to the incident
field at the same values of large detuning [panels (a)-(c)] and near
zero detuning [panels (d)-(e)] inspected in Figs.~\ref{fig3}(c)
and~\ref{fig3}(d). Figures~\ref{fig4}(a) and ~\ref{fig4}(d)
correspond to the high energy extinction peak, Figs.~\ref{fig4}(b)
and ~\ref{fig4}(e) correspond to the extinction dip, and Figs. 4(c)
and 4(f) correspond to the low energy extinction peak.  At large
detuning the field is highly concentrated in the YAG:Ce layer for
(a) - the guided mode -, whereas it is more concentrated near the
metallic structure for (c) - the LSPP. Figure~\ref{fig4}(a) shows
that the guided mode resides primarily in the YAG:Ce waveguide, with
only a small fraction of the total field enhancement in the
Si$_3$N$_4$ layer. Near zero detuning, the field distributions in
(d) and (f) are very similar, since the high and low energy
extinction peaks correspond to hybrid states with approximately
equal weights.  Figure~\ref{fig4}(e) shows strong local field
enhancements in the YAG:Ce layer  at an energy and wave vector for
which the structure is nearly transparent. Comparing
Fig.~\ref{fig4}(e) to 4(b) reveals the critical role of the
eigenmode detuning in the average field enhancement inside the
waveguide at the energy of the extinction minimum. Whereas only a
small enhancement occurs for large detunings, a much larger
enhancement occurs near zero detuning, resulting in a  more intense
emission at the latter condition.

\begin{figure}
\centerline{\scalebox{0.85}{\includegraphics{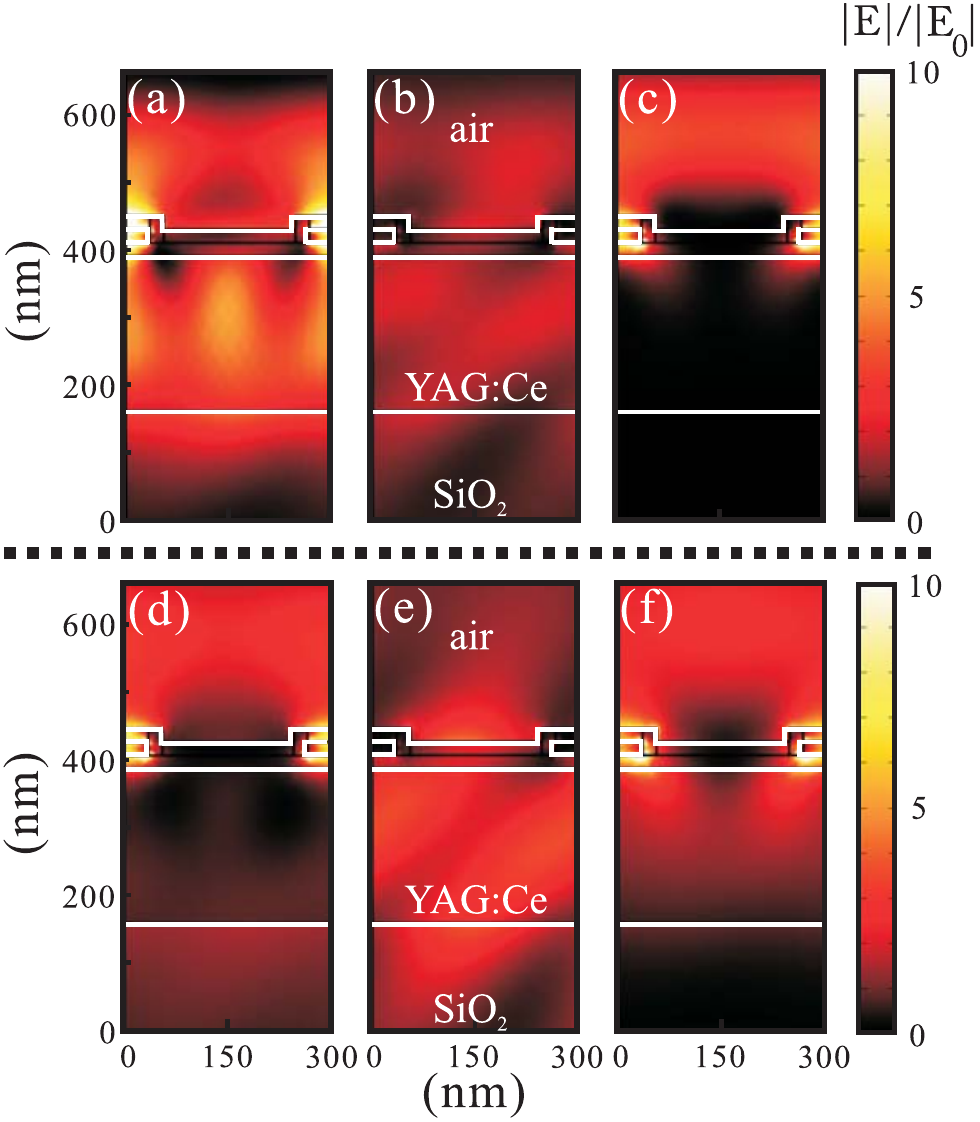}}}
\caption{ Field enhancement for large waveguide-plasmon detuning
(a)-(c), i.e., at $k_{||}=2.1$ mrad/nm, and near zero detuning
(d)-(f), i.e., at $k_{||}=5.5$ mrad/nm.  (a) and (d) correspond to
the high energy extinction peak, (b) and (e) correspond to the
extinction dip, and (c) and (f) correspond to the low energy
extinction peak, all referring to the spectra in Fig. 3(c). The
metallic antennas, surrounded by 20 nm of Si$_3$N$_4$, are located
at the interface between the YAG:Ce layer and air. }\label{fig4}
\end{figure}

\begin{figure}
\centerline{\scalebox{0.63}{\includegraphics{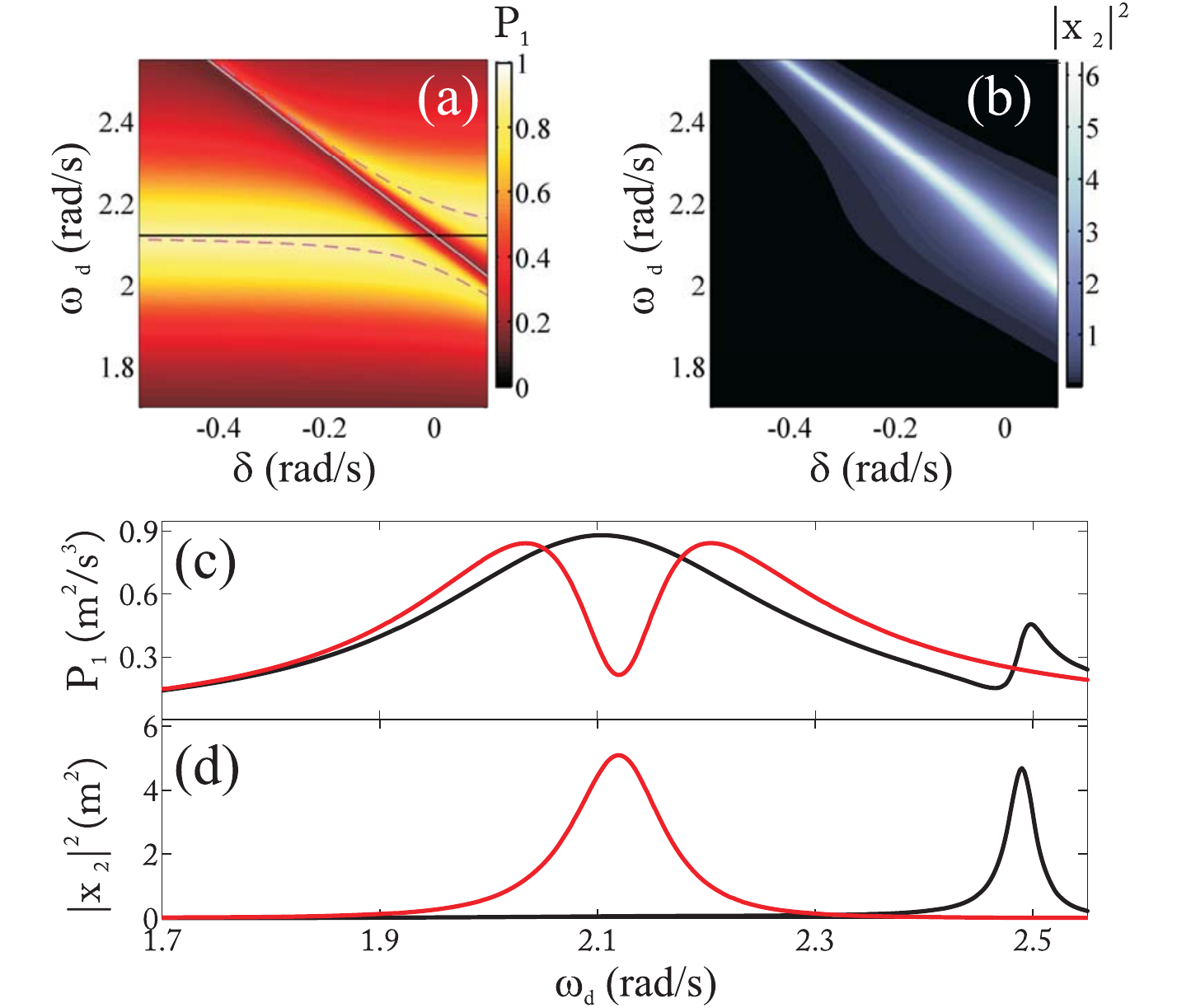}}}
 \caption{(a)
Power dissipated by the first oscillator, and (b) amplitude squared response of the second oscillator, both calculated from Eq. 1. (c) and (d) are cuts of (a) and
(b), respectively, at $\delta=-0.36$ rad/s (black lines) and $\delta=0$ (red lines). In (a) the black solid line represents the eigenfrequency of the first
uncoupled oscillator, the solid cyan line corresponds to the second uncoupled oscillator, and the magenta dashed lines are $\omega_{\pm}$ calculated from Eq.
2.}\label{fig5}
\end{figure}

We elucidate the physics involved in the LSPP-guided-mode coupling
through a model of coupled harmonic oscillators. Classical analogs
to EIT~\cite{Alzar05}, Fano resonances~\cite{Halas10}, and strong
coupling~\cite{Novotny10a}, have been presented with similar models.
The model consists of two harmonic oscillators, one with
eigenfrequency $\omega_0$, and the second with an eigenfrequency
that is a function of the detuning $\delta$ from the eigenfrequency
of the first oscillator, i.e., $\omega_0 - \delta$. Assuming
time-harmonic solutions, the coupled oscillator Hamiltonian is,

\begin{equation}
\left( \begin{matrix}
\omega_0^2 - \omega^2  - i \gamma_1 \omega  & -\Omega_{12}^2  \\
-\Omega_{12}^2 & (\omega_0 - \delta)^2 - \omega^2  - i \gamma_2 \omega ,   
\end{matrix} \right),
\end{equation}

where $\gamma_{1,2}$ are the damping frequencies and $\Omega_{12}$
is the coupling frequency. In analogy to the extinction
measurements, we drive the first oscillator directly by a harmonic
force $F = F_0 e^{-i\omega_d t}$, and calculate the real part of its
dissipated power given by $P_1 = \frac{1}{2}$Re\{$F^* \dot{x_1}$\},
with $\dot{x_1}$ the velocity. We consider that only the first
oscillator is driven directly by the harmonic force $F$ because only
the LSPP is driven directly by the incident electromagnetic field.
Since the dispersion relation of the guided mode lies below the
light, it can only be excited through the antennas.  Finally, we
integrate the dissipated power over one period of oscillation and
scan the driving frequency $\omega_d$ to obtain the dissipated power
spectrum, which we plot as a function of $\delta$ in
Fig.~\ref{fig5}(a). We set the eigenfrequency of the first
oscillator to $\omega_0 = 2.12$ rad/s, the damping frequencies to
$\gamma_1 = 0.4$ rad/s, $\gamma_2 = 0.02$ rad/s, the coupling
frequency to $\Omega_{12} = 0.58$ rad/s, and the force per unit mass
to $F_0 = 0.84$ m/s$^2$.  For the first oscillator we have  $Q_1 =
\omega_0 / \gamma_1 = 5$, which is typical for damped
LSPPs~\cite{Maier}. The lower damping of the second oscillator
reflects the lower losses of the waveguide, wherein absorption is
nearly negligible and out-of-plane scattering losses damp the mode
relatively weakly. The coupling frequency and driving force
amplitude were chosen to match the experiments.

The eigenfrequencies of the uncoupled and coupled oscillators are shown as solid and dashed
lines in Fig.~\ref{fig5}(a), respectively. The eigenfrequencies of the coupled system,

\begin{eqnarray}
\omega_{\pm}^2 &=& \frac{1}{2} [  \omega_0^2 + (\omega_0 - \delta)^2
\pm \\  \nonumber &&  \sqrt{ [\omega_0^2 + (\omega_0 - \delta)^2]^2
- 4 (\omega_0^2 (\omega_0 - \delta)^2 - \Omega_{12}^4 ) } ],
\end{eqnarray}

are obtained from Eq. 1 by letting $\gamma_{1,2} = 0$ and solving for the characteristic equation.
$\omega_{\pm}$ display an avoided crossing, in analogy to the
quantum mechanical dressed states~\cite{Cohen}.
Their energy splitting at zero detuning is the classical counterpart of the Rabi splitting. Notice that for large detunings,
$\omega_{\pm}$ asymptotically approach one of the eigenfrequencies of the uncoupled oscillators, i.e., the bare states.
As $\omega_{\pm}$ each transit through $\delta = 0$, they interchange their resemblance to one or the other of the uncoupled states.
 At $\delta=0$, $\omega_{\pm}$ can not be associated with
either one of the uncoupled oscillators, since the system resides in a hybrid (dressed) state of equal weights.

To model the local field intensity enhancement $|E|^2 / |E_0|^2$ in
the waveguide, we calculate the amplitude squared response of the
associated oscillator, i.e., $|x_2|^2$. It was recently shown that
the amplitude resonance of a single harmonic oscillator may be
associated with the near-field resonance of a metallic
nanoparticle~\cite{Zuloaga11}. Moreover, the shift in the amplitude
response with respect to the dissipated power accounts for the shift
of the near-field with respect to the far-field~\cite{Zuloaga11}. We
herein extend this principle to a coupled system, and show in
Fig.~\ref{fig5}(b) the dependence of this near-field analogous
quantity on the detuning parameter. In Figs.~\ref{fig5}(c) and
~\ref{fig5}(d) we make cuts of both ~\ref{fig5}(a) and
~\ref{fig5}(b) at a large detuning  of $\delta=-0.36$ rad/s (black
line) and at zero detuning (red line). In its simplicity, the
coupled oscillator model conveys the core of the physics involved.
When the eigenfrequencies of the uncoupled oscillators are largely
detuned, the dissipated power spectrum exhibits a Fano-like line
shape near the eigenfrequency of the second oscillator. The enhanced
amplitude response of the second oscillator near this feature
explains the large emission enhancements observed near the guided
mode resonance for large detunings. Moreover, the model shows that
at zero detuning the coupled system displays a minimum in the
dissipated power by one of its constituents (the first oscillator
representing the driven electrons in the metal), while the amplitude
response of the other constituent (the second oscillator
representing the guided mode) is enhanced.

In conclusion, we have demonstrated the strong coupling of localized
surface plasmon polaritons to a guided mode in a light emitting
slab. The properties of the emerging light-emitting
waveguide-plasmon polaritons were shown to be strongly dependent on
the relative detuning between the two eigenmodes involved. Near zero
detuning, light emission in the waveguide is resonantly enhanced in
a strongly dispersive spectral region which is nearly transparent to
the far-field. Our results open a new avenue for investigating
EIT-related phenomena, i.e., slow light, strong dispersion, and
coupling-induced transparencies, in the context of light emission.

 This work was supported by the Netherlands Foundation for
Fundamental Research on Matter (FOM) and the Netherlands
Organisation for Scientific Research (NWO), and is part of an
industrial partnership program between Philips and FOM. We thank G.
Pirruccio for assistance in the simulations. S. Murai acknowledges
financial support from Young Researchers Overseas Visits Program for
Vitalizing Brain Circulation of JSPS, Japan.


\begin{thebibliography}{31}
\expandafter\ifx\csname
natexlab\endcsname\relax\def\natexlab#1{#1}\fi
\expandafter\ifx\csname bibnamefont\endcsname\relax
  \def\bibnamefont#1{#1}\fi
\expandafter\ifx\csname bibfnamefont\endcsname\relax
  \def\bibfnamefont#1{#1}\fi
\expandafter\ifx\csname citenamefont\endcsname\relax
  \def\citenamefont#1{#1}\fi
\expandafter\ifx\csname url\endcsname\relax
  \def\url#1{\texttt{#1}}\fi
\expandafter\ifx\csname urlprefix\endcsname\relax\def\urlprefix{URL
}\fi \providecommand{\bibinfo}[2]{#2}
\providecommand{\eprint}[2][]{\url{#2}}

\bibitem[{\citenamefont{Fano}(1961)}]{Fano}
\bibinfo{author}{\bibfnamefont{U.}~\bibnamefont{Fano}}, \bibinfo{journal}{Phys.
  Rev.} \textbf{\bibinfo{volume}{124}}, \bibinfo{pages}{1866}
  (\bibinfo{year}{1961}).

\bibitem[{\citenamefont{Boller et~al.}(1991)\citenamefont{Boller, Imamoglu, and
  Harris}}]{Harris91}
\bibinfo{author}{\bibfnamefont{K.-J.} \bibnamefont{Boller}},
  \bibinfo{author}{\bibfnamefont{A.}~\bibnamefont{Imamoglu}}, \bibnamefont{and}
  \bibinfo{author}{\bibfnamefont{S.~E.} \bibnamefont{Harris}},
  \bibinfo{journal}{Phys. Rev. Lett.} \textbf{\bibinfo{volume}{66}},
  \bibinfo{pages}{2593} (\bibinfo{year}{1991}).

\bibitem[{\citenamefont{Miroshnichenko
  et~al.}(2010)\citenamefont{Miroshnichenko, Flach, and Kivshar}}]{FanoRev10}
\bibinfo{author}{\bibfnamefont{A.~E.} \bibnamefont{Miroshnichenko}},
  \bibinfo{author}{\bibfnamefont{S.}~\bibnamefont{Flach}}, \bibnamefont{and}
  \bibinfo{author}{\bibfnamefont{Y.~S.} \bibnamefont{Kivshar}},
  \bibinfo{journal}{Rev. Mod. Phys.} \textbf{\bibinfo{volume}{82}},
  \bibinfo{pages}{2257} (\bibinfo{year}{2010}).

\bibitem[{\citenamefont{Luk'yanchuk et~al.}(2010)\citenamefont{Luk'yanchuk,
  Zheludev, Maier, Halas, Nordlander, Giessen, and Chong}}]{FanoNat}
\bibinfo{author}{\bibfnamefont{B.}~\bibnamefont{Luk'yanchuk}},
  \bibinfo{author}{\bibfnamefont{N.~I.} \bibnamefont{Zheludev}},
  \bibinfo{author}{\bibfnamefont{S.~A.} \bibnamefont{Maier}},
  \bibinfo{author}{\bibfnamefont{N.~J.} \bibnamefont{Halas}},
  \bibinfo{author}{\bibfnamefont{P.}~\bibnamefont{Nordlander}},
  \bibinfo{author}{\bibfnamefont{H.}~\bibnamefont{Giessen}}, \bibnamefont{and}
  \bibinfo{author}{\bibfnamefont{C.~T.} \bibnamefont{Chong}},
  \bibinfo{journal}{Nat. Mat.} \textbf{\bibinfo{volume}{9}},
  \bibinfo{pages}{707} (\bibinfo{year}{2010}).

\bibitem[{\citenamefont{Marangos}(1998)}]{Marangos}
\bibinfo{author}{\bibfnamefont{J.~P.} \bibnamefont{Marangos}},
  \bibinfo{journal}{Journal of Modern Optics} \textbf{\bibinfo{volume}{45}},
  \bibinfo{pages}{471} (\bibinfo{year}{1998}).

\bibitem[{\citenamefont{Fleischhauer et~al.}(2005)\citenamefont{Fleischhauer,
  Imamoglu, and Marangos}}]{Fleis}
\bibinfo{author}{\bibfnamefont{M.}~\bibnamefont{Fleischhauer}},
  \bibinfo{author}{\bibfnamefont{A.}~\bibnamefont{Imamoglu}}, \bibnamefont{and}
  \bibinfo{author}{\bibfnamefont{J.~P.} \bibnamefont{Marangos}},
  \bibinfo{journal}{Rev. Mod. Phys.} \textbf{\bibinfo{volume}{77}},
  \bibinfo{pages}{633} (\bibinfo{year}{2005}).

\bibitem[{\citenamefont{Zhang et~al.}(2008)\citenamefont{Zhang, Genov, Wang,
  Liu, and Zhang}}]{Zhang08}
\bibinfo{author}{\bibfnamefont{S.}~\bibnamefont{Zhang}},
  \bibinfo{author}{\bibfnamefont{D.~A.} \bibnamefont{Genov}},
  \bibinfo{author}{\bibfnamefont{Y.}~\bibnamefont{Wang}},
  \bibinfo{author}{\bibfnamefont{M.}~\bibnamefont{Liu}}, \bibnamefont{and}
  \bibinfo{author}{\bibfnamefont{X.}~\bibnamefont{Zhang}},
  \bibinfo{journal}{Phys. Rev. Lett.} \textbf{\bibinfo{volume}{101}},
  \bibinfo{pages}{047401} (\bibinfo{year}{2008}).

\bibitem[{\citenamefont{Liu et~al.}(2009)\citenamefont{Liu, Langguth, Weiss,
  K\"{a}stel, Fleischhauer, Pfau, and Giessen}}]{Giessen09}
\bibinfo{author}{\bibfnamefont{N.}~\bibnamefont{Liu}},
  \bibinfo{author}{\bibfnamefont{L.}~\bibnamefont{Langguth}},
  \bibinfo{author}{\bibfnamefont{T.}~\bibnamefont{Weiss}},
  \bibinfo{author}{\bibfnamefont{J.}~\bibnamefont{K\"{a}stel}},
  \bibinfo{author}{\bibfnamefont{M.}~\bibnamefont{Fleischhauer}},
  \bibinfo{author}{\bibfnamefont{T.}~\bibnamefont{Pfau}}, \bibnamefont{and}
  \bibinfo{author}{\bibfnamefont{H.}~\bibnamefont{Giessen}},
  \bibinfo{journal}{Nat. Mat.} \textbf{\bibinfo{volume}{8}},
  \bibinfo{pages}{758} (\bibinfo{year}{2009}).

\bibitem[{\citenamefont{Yannopapas et~al.}(2009)\citenamefont{Yannopapas,
  Paspalakis, and Vitanov}}]{Yannopapas09}
\bibinfo{author}{\bibfnamefont{V.}~\bibnamefont{Yannopapas}},
  \bibinfo{author}{\bibfnamefont{E.}~\bibnamefont{Paspalakis}},
  \bibnamefont{and} \bibinfo{author}{\bibfnamefont{N.~V.}
  \bibnamefont{Vitanov}}, \bibinfo{journal}{Phys. Rev. B}
  \textbf{\bibinfo{volume}{80}}, \bibinfo{pages}{035104}
  (\bibinfo{year}{2009}).

\bibitem[{\citenamefont{Tassin et~al.}(2009)\citenamefont{Tassin, Zhang,
  Koschny, Economou, and Soukoulis}}]{Soukoulis09}
\bibinfo{author}{\bibfnamefont{P.}~\bibnamefont{Tassin}},
  \bibinfo{author}{\bibfnamefont{L.}~\bibnamefont{Zhang}},
  \bibinfo{author}{\bibfnamefont{T.}~\bibnamefont{Koschny}},
  \bibinfo{author}{\bibfnamefont{E.~N.} \bibnamefont{Economou}},
  \bibnamefont{and} \bibinfo{author}{\bibfnamefont{C.~M.}
  \bibnamefont{Soukoulis}}, \bibinfo{journal}{Phys. Rev. Lett.}
  \textbf{\bibinfo{volume}{102}}, \bibinfo{pages}{053901}
  (\bibinfo{year}{2009}).

\bibitem[{\citenamefont{Kekatpure et~al.}(2010)\citenamefont{Kekatpure,
  Barnard, Cai, and Brongersma}}]{Kekatpure&Brongersma10}
\bibinfo{author}{\bibfnamefont{R.~D.} \bibnamefont{Kekatpure}},
  \bibinfo{author}{\bibfnamefont{E.~S.} \bibnamefont{Barnard}},
  \bibinfo{author}{\bibfnamefont{W.}~\bibnamefont{Cai}}, \bibnamefont{and}
  \bibinfo{author}{\bibfnamefont{M.~L.} \bibnamefont{Brongersma}},
  \bibinfo{journal}{Phys. Rev. Lett.} \textbf{\bibinfo{volume}{104}},
  \bibinfo{pages}{243902} (\bibinfo{year}{2010}).

\bibitem[{\citenamefont{Zhang et~al.}(2011)\citenamefont{Zhang, Bai, Cai, Xu,
  Song, and Gan}}]{Zhang11APL}
\bibinfo{author}{\bibfnamefont{J.}~\bibnamefont{Zhang}},
  \bibinfo{author}{\bibfnamefont{W.}~\bibnamefont{Bai}},
  \bibinfo{author}{\bibfnamefont{L.}~\bibnamefont{Cai}},
  \bibinfo{author}{\bibfnamefont{Y.}~\bibnamefont{Xu}},
  \bibinfo{author}{\bibfnamefont{G.}~\bibnamefont{Song}}, \bibnamefont{and}
  \bibinfo{author}{\bibfnamefont{Q.}~\bibnamefont{Gan}},
  \bibinfo{journal}{Appl. Phys. Lett.} \textbf{\bibinfo{volume}{99}},
  \bibinfo{pages}{181120} (\bibinfo{year}{2011}).


\bibitem[{\citenamefont{Hau et~al.}(1999)\citenamefont{Hau, Harris, Dutton, and
  Behroozi}}]{Hau99}
\bibinfo{author}{\bibfnamefont{L.}~\bibnamefont{Hau}},
  \bibinfo{author}{\bibfnamefont{S.~E.} \bibnamefont{Harris}},
  \bibinfo{author}{\bibfnamefont{Z.}~\bibnamefont{Dutton}}, \bibnamefont{and}
  \bibinfo{author}{\bibfnamefont{C.~H.} \bibnamefont{Behroozi}},
  \bibinfo{journal}{Nature} \textbf{\bibinfo{volume}{397}},
  \bibinfo{pages}{584} (\bibinfo{year}{1999}).

\bibitem[{\citenamefont{Liu et~al.}(2001)\citenamefont{Liu, Dutton, Behroozi,
  and Hau}}]{Hau01}
\bibinfo{author}{\bibfnamefont{C.}~\bibnamefont{Liu}},
  \bibinfo{author}{\bibfnamefont{Z.}~\bibnamefont{Dutton}},
  \bibinfo{author}{\bibfnamefont{C.~H.} \bibnamefont{Behroozi}},
  \bibnamefont{and} \bibinfo{author}{\bibfnamefont{L.~V.} \bibnamefont{Hau}},
  \bibinfo{journal}{Nature} \textbf{\bibinfo{volume}{409}},
  \bibinfo{pages}{490} (\bibinfo{year}{2001}).

\bibitem[{\citenamefont{Bouwmeester et~al.}(1995)\citenamefont{Bouwmeester,
  Dekker, Dorsselaer, Schrama, Visser, and Woerdman}}]{Woerdman}
\bibinfo{author}{\bibfnamefont{D.}~\bibnamefont{Bouwmeester}},
  \bibinfo{author}{\bibfnamefont{N.~H.} \bibnamefont{Dekker}},
  \bibinfo{author}{\bibfnamefont{F.~E.~v.} \bibnamefont{Dorsselaer}},
  \bibinfo{author}{\bibfnamefont{C.~A.} \bibnamefont{Schrama}},
  \bibinfo{author}{\bibfnamefont{P.~M.} \bibnamefont{Visser}},
  \bibnamefont{and} \bibinfo{author}{\bibfnamefont{J.~P.}
  \bibnamefont{Woerdman}}, \bibinfo{journal}{Phys. Rev. A}
  \textbf{\bibinfo{volume}{51}}, \bibinfo{pages}{646} (\bibinfo{year}{1995}).

\bibitem[{\citenamefont{Novotny}(2010)}]{Novotny10a}
\bibinfo{author}{\bibfnamefont{L.}~\bibnamefont{Novotny}},
  \bibinfo{journal}{Am. J. Phys.} \textbf{\bibinfo{volume}{78}},
  \bibinfo{pages}{1199} (\bibinfo{year}{2010}).

\bibitem[{\citenamefont{M\"uhlschlegel
  et~al.}(2005)\citenamefont{M\"uhlschlegel, Eisler, Martin, Hecht, and
  Pohl}}]{Hecht05}
\bibinfo{author}{\bibfnamefont{P.}~\bibnamefont{M\"uhlschlegel}},
  \bibinfo{author}{\bibfnamefont{H.-J.} \bibnamefont{Eisler}},
  \bibinfo{author}{\bibfnamefont{O.~J.~F.} \bibnamefont{Martin}},
  \bibinfo{author}{\bibfnamefont{B.}~\bibnamefont{Hecht}}, \bibnamefont{and}
  \bibinfo{author}{\bibfnamefont{D.~W.} \bibnamefont{Pohl}},
  \bibinfo{journal}{Science} \textbf{\bibinfo{volume}{308}},
  \bibinfo{pages}{1607} (\bibinfo{year}{2005}).

\bibitem[{\citenamefont{Novotny and van Hulst}(2011)}]{Novotny11}
\bibinfo{author}{\bibfnamefont{L.}~\bibnamefont{Novotny}} \bibnamefont{and}
  \bibinfo{author}{\bibfnamefont{N.}~\bibnamefont{van Hulst}},
  \bibinfo{journal}{Nat. Phot.} \textbf{\bibinfo{volume}{5}},
  \bibinfo{pages}{83} (\bibinfo{year}{2011}).

\bibitem[{\citenamefont{Christ et~al.}(2003)\citenamefont{Christ, Tikhodeev,
  Gippius, Kuhl, and Giessen}}]{Giessen03}
\bibinfo{author}{\bibfnamefont{A.}~\bibnamefont{Christ}},
  \bibinfo{author}{\bibfnamefont{S.~G.} \bibnamefont{Tikhodeev}},
  \bibinfo{author}{\bibfnamefont{N.~A.} \bibnamefont{Gippius}},
  \bibinfo{author}{\bibfnamefont{J.}~\bibnamefont{Kuhl}}, \bibnamefont{and}
  \bibinfo{author}{\bibfnamefont{H.}~\bibnamefont{Giessen}},
  \bibinfo{journal}{Phys. Rev. Lett.} \textbf{\bibinfo{volume}{91}},
  \bibinfo{pages}{183901} (\bibinfo{year}{2003}).

\bibitem[{\citenamefont{Murai et~al.}(2012)\citenamefont{Murai, Verschuuren,
  Lozano, Pirruccio, Koenderink, and Gomez~Rivas}}]{Murai}
\bibinfo{author}{\bibfnamefont{S.}~\bibnamefont{Murai}},
  \bibinfo{author}{\bibfnamefont{M.~A.} \bibnamefont{Verschuuren}},
  \bibinfo{author}{\bibfnamefont{G.}~\bibnamefont{Lozano}},
  \bibinfo{author}{\bibfnamefont{G.}~\bibnamefont{Pirruccio}},
  \bibinfo{author}{\bibfnamefont{A.~F.} \bibnamefont{Koenderink}},
  \bibnamefont{and}
  \bibinfo{author}{\bibfnamefont{J.}~\bibnamefont{Gomez~Rivas}},
  \bibinfo{journal}{Optical Materials Express} \textbf{\bibinfo{volume}{2}},
  \bibinfo{pages}{1111} (\bibinfo{year}{2012}).

\bibitem[{\citenamefont{Wokaun et~al.}(1983)\citenamefont{Wokaun, Lutz, King,
  Wild, and Ernst}}]{Wokaun83}
\bibinfo{author}{\bibfnamefont{A.}~\bibnamefont{Wokaun}},
  \bibinfo{author}{\bibfnamefont{H.-P.} \bibnamefont{Lutz}},
  \bibinfo{author}{\bibfnamefont{A.~P.} \bibnamefont{King}},
  \bibinfo{author}{\bibfnamefont{U.~P.} \bibnamefont{Wild}}, \bibnamefont{and}
  \bibinfo{author}{\bibfnamefont{R.~R.} \bibnamefont{Ernst}},
  \bibinfo{journal}{J. Chem. Phys.} \textbf{\bibinfo{volume}{79}},
  \bibinfo{pages}{509} (\bibinfo{year}{1983}).

\bibitem[{\citenamefont{Anger et~al.}(2006)\citenamefont{Anger, Bharadwaj, and
  Novotny}}]{Anger&Novotny06}
\bibinfo{author}{\bibfnamefont{P.}~\bibnamefont{Anger}},
  \bibinfo{author}{\bibfnamefont{P.}~\bibnamefont{Bharadwaj}},
  \bibnamefont{and} \bibinfo{author}{\bibfnamefont{L.}~\bibnamefont{Novotny}},
  \bibinfo{journal}{Phys. Rev. Lett.} \textbf{\bibinfo{volume}{96}},
  \bibinfo{pages}{113002} (\bibinfo{year}{2006}).

\bibitem[{\citenamefont{Verschuuren}(2010)}]{SCIL}
\bibinfo{author}{\bibfnamefont{M.~A.} \bibnamefont{Verschuuren}},
  \bibinfo{type}{{PhD} dissertation}, \bibinfo{school}{Utrecht University}
  (\bibinfo{year}{2010}).

\bibitem[{\citenamefont{Palik}(1985)}]{Palik91}
\bibinfo{author}{\bibfnamefont{E.}~\bibnamefont{Palik}},
  \emph{\bibinfo{title}{Handbook of optical constants of solids}}
  (\bibinfo{publisher}{Academic Press}, \bibinfo{address}{New York},
  \bibinfo{year}{1985}).

\bibitem[{\citenamefont{Yariv and Yeh}(2007)}]{Yariv}
\bibinfo{author}{\bibfnamefont{A.}~\bibnamefont{Yariv}} \bibnamefont{and}
  \bibinfo{author}{\bibfnamefont{P.}~\bibnamefont{Yeh}},
  \emph{\bibinfo{title}{Photonics: Optical Electronics in Modern
  Communications}} (\bibinfo{publisher}{Oxford University Press},
  \bibinfo{address}{Oxford}, \bibinfo{year}{2007}), \bibinfo{edition}{6th} ed.

\bibitem[{\citenamefont{Al\`u and Engheta}(2009)}]{Alu09}
\bibinfo{author}{\bibfnamefont{A.}~\bibnamefont{Al\`u}} \bibnamefont{and}
  \bibinfo{author}{\bibfnamefont{N.}~\bibnamefont{Engheta}},
  \bibinfo{journal}{Phys. Rev. Lett.} \textbf{\bibinfo{volume}{102}},
  \bibinfo{pages}{233901} (\bibinfo{year}{2009}).

\bibitem[{\citenamefont{de~Abajo}(2009)}]{Garcia09}
\bibinfo{author}{\bibfnamefont{F.~J.~G.} \bibnamefont{de~Abajo}},
  \bibinfo{journal}{Physics} \textbf{\bibinfo{volume}{2}}, \bibinfo{pages}{47}
  (\bibinfo{year}{2009}).

\bibitem[{\citenamefont{Alzar et~al.}(2002)\citenamefont{Alzar, Martinez, and
  Nussenzveig}}]{Alzar05}
\bibinfo{author}{\bibfnamefont{C.}~\bibnamefont{Alzar}},
  \bibinfo{author}{\bibfnamefont{M.}~\bibnamefont{Martinez}}, \bibnamefont{and}
  \bibinfo{author}{\bibfnamefont{P.}~\bibnamefont{Nussenzveig}},
  \bibinfo{journal}{Am. J. Phys.} \textbf{\bibinfo{volume}{70}},
  \bibinfo{pages}{37} (\bibinfo{year}{2002}).

\bibitem[{\citenamefont{Mukherjee et~al.}(2010)\citenamefont{Mukherjee,
  Sobhani, Lassiter, Bardhan, Nordlander, and Halas}}]{Halas10}
\bibinfo{author}{\bibfnamefont{S.}~\bibnamefont{Mukherjee}},
  \bibinfo{author}{\bibfnamefont{H.}~\bibnamefont{Sobhani}},
  \bibinfo{author}{\bibfnamefont{J.~B.} \bibnamefont{Lassiter}},
  \bibinfo{author}{\bibfnamefont{R.}~\bibnamefont{Bardhan}},
  \bibinfo{author}{\bibfnamefont{P.}~\bibnamefont{Nordlander}},
  \bibnamefont{and} \bibinfo{author}{\bibfnamefont{N.~J.} \bibnamefont{Halas}},
  \bibinfo{journal}{Nano Lett.} \textbf{\bibinfo{volume}{10}},
  \bibinfo{pages}{2694} (\bibinfo{year}{2010}).


\bibitem[{\citenamefont{Maier}(2007)}]{Maier}
\bibinfo{author}{\bibfnamefont{S.~A.} \bibnamefont{Maier}},
  \emph{\bibinfo{title}{Plasmonics: Fundamentals and Applications}}
  (\bibinfo{publisher}{Springer}, \bibinfo{address}{New York, USA},
  \bibinfo{year}{2007}).

\bibitem[{\citenamefont{Cohen-Tannoudji
  et~al.}(2004)\citenamefont{Cohen-Tannoudji, Dupont-Roc, and
  Grynberg}}]{Cohen}
\bibinfo{author}{\bibnamefont{Cohen-Tannoudji}},
  \bibinfo{author}{\bibfnamefont{C.~J.} \bibnamefont{Dupont-Roc}},
  \bibnamefont{and} \bibinfo{author}{\bibfnamefont{G.}~\bibnamefont{Grynberg}},
  \emph{\bibinfo{title}{Atom - Photon Interactions}}
  (\bibinfo{publisher}{Wiley-VCH}, \bibinfo{address}{{N}ew {Y}ork},
  \bibinfo{year}{2004}).

\bibitem[{\citenamefont{Zuloaga and Nordlander}(2011)}]{Zuloaga11}
\bibinfo{author}{\bibfnamefont{J.}~\bibnamefont{Zuloaga}} \bibnamefont{and}
  \bibinfo{author}{\bibfnamefont{P.}~\bibnamefont{Nordlander}},
  \bibinfo{journal}{Nano Letters} \textbf{\bibinfo{volume}{11}},
  \bibinfo{pages}{1280} (\bibinfo{year}{2011}).

\end{thebibliography}

\end{document}